\documentclass[aps,prl,twocolumn,amsmath,amssymb,nofootinbib,superscriptaddress,floatfix,reprint,longbibliography]{revtex4-1}
\usepackage[dvips]{graphicx}
\usepackage{latexsym}
\usepackage{amsmath}
\usepackage{amsfonts}
\usepackage{amssymb}
\usepackage{bm}
\usepackage{color}
\usepackage{txfonts}
\usepackage{float}
\usepackage{url}
\usepackage[colorlinks=true, urlcolor=blue, linkcolor=blue, citecolor=blue, pdftex]{hyperref}
\usepackage{ulem}
\usepackage{physics}
\usepackage{pifont}
\usepackage{lineno}
\begin{document}
	\newcommand{\fig}[2]{\includegraphics[width=#1]{#2}}
	\newcommand{\pprl}{Phys. Rev. Lett. \ }
	\newcommand{\pprb}{Phys. Rev. {B}}

\title {Bound states in doped charge transfer insulators}
\author{Pengfei Li}
\thanks{These authors contributed equally to this work.}
\affiliation{Beijing National Laboratory for Condensed Matter Physics and Institute of Physics,
	Chinese Academy of Sciences, Beijing 100190, China}
\affiliation{School of Physical Sciences, University of Chinese Academy of Sciences, Beijing 100190, China}
\affiliation{Department of Physics, The Hong Kong University of Science and Technology, Clear Water Bay, Kowloon, Hong Kong, China}

\author{Yang Shen}
\thanks{These authors contributed equally to this work.}
\affiliation{Key Laboratory of Artificial Structures and Quantum Control (Ministry of Education), School of Physics and Astronomy, Shanghai Jiao Tong University, Shanghai 200240, China}

\author{Mingpu Qin}
\thanks{Corresponding author: qinmingpu@sjtu.edu.cn}
\affiliation{Key Laboratory of Artificial Structures and Quantum Control (Ministry of Education), School of Physics and Astronomy, Shanghai Jiao Tong University, Shanghai 200240, China}

\author{Kun Jiang}
\thanks{Corresponding author: jiangkun@iphy.ac.cn}
\affiliation{Beijing National Laboratory for Condensed Matter Physics and Institute of Physics,
	Chinese Academy of Sciences, Beijing 100190, China}
\affiliation{School of Physical Sciences, University of Chinese Academy of Sciences, Beijing 100190, China}

\author{Jiangping Hu}
\affiliation{Beijing National Laboratory for Condensed Matter Physics and Institute of Physics,
	Chinese Academy of Sciences, Beijing 100190, China}
\affiliation{School of Physical Sciences, University of Chinese Academy of Sciences, Beijing 100190, China}
\affiliation{New Cornerstone Science Laboratory, Institute of Physics, Chinese Academy of Sciences, Beijing 100190, China}

\author{Tao Xiang}
\affiliation{Beijing National Laboratory for Condensed Matter Physics and Institute of Physics,
	Chinese Academy of Sciences, Beijing 100190, China}
\affiliation{School of Physical Sciences, University of Chinese Academy of Sciences, Beijing 100190, China}

\date{\today}

\begin{abstract}
Understanding the physics of doped charge transfer insulators is the most important problem in high-temperature superconductivity. In this work, we show that an in-gap bound state emerges from the localized hole of the doped charge transfer insulator. We propose an approximate ground state wavefunction based on one localized Zhang-Rice singlet and the Neel state. By calculating the excitation states with one hole added and removed from this ground state, we successfully identify the existence of bound states inside the charge transfer gap. This feature is further confirmed by a Lanczos calculation based on matrix product states (MPS) for a system of $4\times4$ CuO$_2$ unit cells. How these bound states evolve into metallic states is further discussed. Our findings identify the key component of recent STM results on lightly doped Ca$_2$CuO$_2$Cl$_2$ and provide a new understanding of hole-doped charge transfer insulators.

\end{abstract}
\maketitle

\section{Introduction}
As a relative of Mott insulator, the concept of charge transfer (CT) insulator plays an important role in the understanding of the insulating behaviors in correlated transition metal insulators, especially in cuprates \cite{Zannen_PhysRevLett.55.418,khomskii2014transition}. The key energy scale for charge transfer insulators is the charge-transfer energy between $d$ electrons and their neighbor $p$ electrons $\Delta_{\mathrm{CT}}$, instead of the onsite Hubbard repulsion energy $U$ for the $d$ electrons. Therefore, the doped holes in cuprates reside primarily on oxygen sites and together with the hole in the Cu site. They form the well-known Zhang-Rice (ZR) singlets because the Cu$^{3+}$ ($3d^8$) state has a higher energy \cite{Zhang_Rice}. These singlets are the dominant charge carriers in the hole-doped cuprates and become superconducting under the influence of super-exchange interaction $J\mathbf{S}_i\cdot\mathbf{S}_j$ between $d$ electrons of Cu \cite{Zhang_Rice}. Although this effective single-band $t$-$J$ model and the corresponding Hubbard model are extensively explored \cite{t-j-dmrg,qin_PhysRevX.10.031016} and the condition for the appearance of superconductivity is studied recently \cite{doi:10.1126/science.adh7691,PhysRevLett.132.066002,PhysRevLett.127.097002}, 
a comprehensive understanding of the process of doping the charge transfer insulator and Mott insulator is a central theme for the ultimate understanding of the high-temperature superconductivity \cite{doping_mott}. 

\begin{figure}[t]
	\begin{center}
		\fig{3.4in}{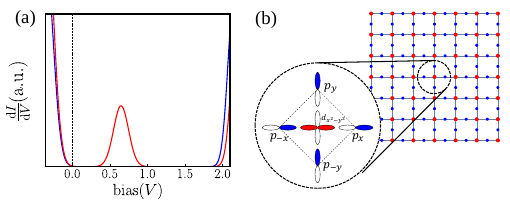}
		\caption{(a) The schematic scanning tunneling spectroscopy density of states in lightly Na doped cuprates Ca$_2$CuO$_2$Cl$_2$ \cite{ye2023visualizing}. The blue line is the spectrum away from the Na-doped center while the red line is the spectrum at the Na-doped center. A bound state around $E_{b} \simeq 0.6-0.7 ~\mathrm{eV}$ shows up at the charge transfer gap.  (b) The essential CuO$_2$ lattice of cuprates superconductors. Around each Cu, there are four neighbor oxygens. The most relevant orbitals are Cu 3$d_{x^2-y^2}$, O 2$p_{x}$ and O 2$p_{y}$,  as illustrated in the inset. 
			\label{fig1}}
	\end{center}
\end{figure}

Recently, a new scanning tunneling microscopy (STM) experiment was carried out on the extremely lightly doped cuprates Ca$_2$CuO$_2$Cl$_2$ (CCOC) \cite{ye2023visualizing}. The Na$^{1+}$ atoms are introduced to replace Ca$^{2+}$ atoms resulting in hole doping. Surprisingly, bound states inside the charge transfer gap of CCOC were observed, as illustrated in Fig.\ref{fig1}(a) \cite{ye2023visualizing}. The spatial tunneling spectrum distribution for these bound states shows a similar contour of the Zhang-Rice singlet.
This new observation provides a great opportunity to understand the hole-doped charge-transfer insulators and Mott insulators
\cite{Sawatzky_PhysRevLett.67.1035,Sawatzky_PhysRevB.48.3916,Dagotto_RevModPhys.66.763,Shen_RevModPhys.75.473,Phillips_RevModPhys.82.1719}.
In this work, we provide a comprehensive study of the doped charge transfer insulator and point out the bound states are indeed related to the localized Zhang-Rice singlets.

There are two important energy scales for the above observations, the CT gap $\Delta_{\mathrm{CT}} \simeq 2 ~\mathrm{eV}$ and the bound state energy $E_{b} \simeq 0.6-0.7 ~\mathrm{eV}$. These two energy scales are much larger than the exchange energy $J\sim 0.14 ~\mathrm{eV}$ and the hopping scale $t\sim 0.35 ~\mathrm{eV}$ \cite{kim_PhysRevLett.80.4245}.
Hence, the kinetic energy and exchange energy can be treated as perturbations for the bound states in a doped CT insulator. This is the central starting point for our following discussions. Another important fact is that the bound states observed in STM experiment \cite{ye2023visualizing} are localized, or more precisely, the ZR singlets are localized. 
Physically, this localization is owing to the poor screening Coulomb potential between Na and the doped hole in the CT insulator.
A similar binding occurs in dilute doped semiconductors\cite{shockley1950electrons,kohn1957shallow}, where the doped electrons or holes are trapped by dopants attraction potentials.
In the following discussion, we will show that the appearance of in-gap bound states results from the localized ZR singlet state.

With these two facts, we start from a qualitative perspective and propose the below classical product state wavefunction $|GS\rangle$ as the approximated ground state wavefunction, which provides a qualitative description of the hole-doped CT insulator in the background of antiferromagnetism (AFM):
\begin{equation}
    |GS\rangle= \psi_{0A}^{\dagger}d_{0A\uparrow}|\mathrm{AFM}\rangle.
    \label{eq:wf}
\end{equation}
Following the conventional notation \cite{Zhang_Rice}, we define the vacuum $|\mathrm{vac}\rangle$ as the fully filled Cu $d^{10}$ and O $p^6$ states from the hole picture. $d_{i\alpha\sigma}^{\dagger}$ is the hole creation operator for Cu 3$d_{x^2-y^2}$ at unit cell $i$ and AFM sublattice index $\alpha=A,B$. 
$|\mathrm{AFM}\rangle=\prod_i d_{iA\uparrow}^{\dagger}d_{iB\downarrow}^{\dagger} \ket{\mathrm{\mathrm{vac}}}$  is the classical N\'eel AFM wavefunction.
The $\psi_{0A}^{\dagger}$ is the ZR singlet operator for the bound site at center 0 and sublattice A. The exact form of $\psi_{0A}^{\dagger}$ will be discussed below as an extension of the ZR singlet proposed in the original Zhang-Rice paper \cite{Zhang_Rice}. We also assume there are total $N$ holes for the ground states. More precisely, there is one hole for each magnetic site and two holes for the ZR site.

Indeed, the classical AFM N\'eel state should not be the ground state for the Heisenberg model and the half-filled Hubbard model owing to the quantum fluctuations in two dimensions \cite{AFM_PhysRev.86.694}. However, the energy difference between the Neel state and the true ground state should be less than order $J$ \cite{AFM_PhysRev.86.694}. Then, we can safely argue the Neel state is enough for the localized bound state problem.
On the other hand, $\psi_{0A}^{\dagger}$ may also deviate from the exact state, which is also a controllable perturbation. 
Therefore, $|GS\rangle$ is a faithful approximate ground state.

Before a more detailed discussion, we introduce the three-band Hamiltonian \cite{Emery_PhysRevLett.58.2794,Zhang_Rice} below
\begin{eqnarray}
H&=&\tilde{\epsilon}_d \sum_{i} d_{i\sigma}^\dagger d_{i\sigma}+\tilde{\epsilon}_p  \sum_{l} p_{l\sigma}^\dagger p_{l\sigma} +U \sum_{i} \hat{n}_{d\uparrow}\hat{n}_{d\downarrow}\nonumber \\
 &+&\sum_{\langle il \rangle} t_{pd}(d_{i\sigma}^\dagger p_{l\sigma}+\mathrm{h.c.})
 -\sum_{\langle ll' \rangle} t_{pp}(p_{l\sigma}^\dagger p_{l'\sigma}+\mathrm{h.c.}) \nonumber\\
 &+& E_{\mathrm{loc}} \sum_{\delta} p_{0\delta\sigma}^\dagger p_{0\delta\sigma} \label{eq:three_band}
\end{eqnarray}
where  $p_{l\sigma}^{\dagger}$ is the hole creation operator for O $(2p_{x},2p_{y})$ at site $l$. $t_{pd}$ and $t_{pp}$ are hopping integrals involving $p-d$ and $p-p$ processes respectively. $\tilde{\epsilon}_{d/p}$ is the on-site potential for $d/p$ orbitals in the hole picture respectively. In the hole representation, the vacuum state is defined as the fully filled Cu $3d^{10}$ and O $2p^{6}$ configuration and the undoped AFM is the configuration with one hole.
More importantly, we add a $E_{\mathrm{loc}}$ term for the local potential energy at O atoms around the bound site due to dopants, which stabilize a localized ZR solution. Owing to poor Coulomb screening, the binding between the doped hole and dopants is much stronger and the bound states are more localized compared to semiconductors.


Since we focus on the tunneling spectrum of STM, the tunneling density of states (DOS) at voltage V at $\mathbf{r}$ following Fermi's golden rule can be written as
\begin{eqnarray}
    \frac{\mathrm{d}I}{\mathrm{d}V}(\mathbf{r},\omega) \propto \sum_{n} |\langle n| \phi_{\sigma}^{\dagger}(\mathbf{r}) |GS\rangle|^2 \delta(-\omega-E_n+E_{\mathrm{GS}}) \nonumber \\
    +\sum_{m} |\langle m| \phi_{\sigma}(\mathbf{r})|GS\rangle|^2 \delta(-\omega-E_{\mathrm{GS}}+E_{m})
    \label{dIdV}
\end{eqnarray} 
where $\phi_{\sigma}^{\dagger}(\mathbf{r})$ will be equal to $ d_{i\sigma}^\dagger$ or  $p_{l\sigma}^\dagger$ depending on the position and $\omega$ is the energy in the unit of $\mathrm{eV}$. $|n\rangle$ are adding hole excitation eigenstates with hole number $n>N$ and $|m\rangle$ are removing hole excitation eigenstates with hole number $m<N$. 
Note that we have switched the energy $\omega$ in the delta functions to the electron picture for experimental comparison.
Hence, the gap for the insulating behavior is related to 
\begin{eqnarray}
    \Delta_{\mathrm{gap}}=E_{N+1}+E_{N-1}-2E_{\mathrm{GS}} 
    \label{eq:charge_gap}
\end{eqnarray}

\begin{figure}[t]
	\begin{center}
		\fig{3.4in}{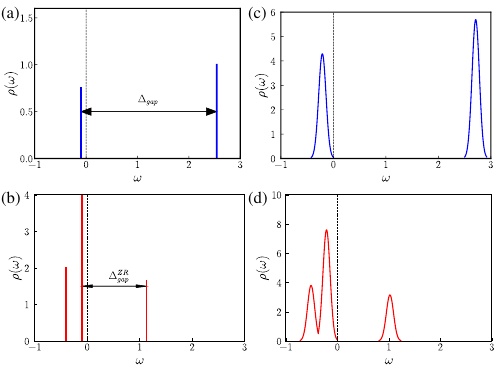}
		\caption{(a),(b) The tunneling DOS $\rho(\omega)$ calculated away from the ZR center and at the ZR center of ground state $|GS\rangle$ respectively. The three-band parameters we used here are adapted from an ab initio calculation for CCOC with $t_{pd}=0.78eV$, $t_{pp}=0.15eV$, $\tilde{\epsilon}_{p}-\tilde{\epsilon}_{d}=3.25eV$, $U=9.5eV$ \cite{kent_PhysRevB.78.035132}. We add the local potential $E_{\mathrm{loc}}=-2t_{pd}$ on O atoms. (c),(d) The schematic broadening tunneling DOS of (a),(b) taking other effects into account, especially the motion effect.
			\label{fig2}}
	\end{center}
\end{figure}

\section{Results}
\subsection{One-cell-calculation results}

Now, we can apply the above formulas to our case. Firstly, we focus on the spectrum away from the ZR center. The tunneling process is just equivalent to the AFM tunneling spectrum. 
For the $|n\rangle$ excitation with $N+1$ holes, the problem is reduced to a single hole moving in the background of the quantum antiferromagnet \cite{varma_PhysRevLett.60.2793,kane_PhysRevB.39.6880}. 
As discussed above, $t$ and $J$ are perturbations. We can use the local eigenstates as the approximated states for the spectrum problem.
For example, we use the localized ZR singlet to replace the moving hole as the approximation excited state $|n\rangle$ \cite{Chernyshev_PhysRevB.50.13768}. 
Hence, the spectrum problem reduces to exact diagonalization problems around one Cu cell as illustrated in Fig.\ref{fig1} (b) \cite{Feiner_PhysRevB.45.7959}. Using the local $D_4$ symmetry, we can group the four $p$ orbital into one $b_1$, one $a_1$ and two degenerate $e$ symmetric basis:
\begin{eqnarray}
    b_{\sigma}&=&(p_{x\sigma}-p_{y\sigma}-p_{-x\sigma}+p_{-y\sigma})/2 \nonumber \\
    a_{\sigma}&=&(p_{x\sigma}+p_{y\sigma}-p_{-x\sigma}-p_{-y\sigma})/2  \label{eq:local_symmetry_main}\\
    e_{1\sigma}&=&(p_{x\sigma}+p_{-x\sigma})/{\sqrt{2}}, e_{2\sigma}=(p_{y\sigma}+p_{-y\sigma})/{\sqrt{2}} \nonumber
\end{eqnarray}
Their corresponding energies become $\epsilon_b=\tilde{\epsilon}_p-2t_{pp}$,$\epsilon_a=\tilde{\epsilon}_p+2t_{pp}$ and $\epsilon_e=\tilde{\epsilon}_p$. Due to $d_\sigma$ only has $b_1$ symmetry, it only has overlap with $b_{\sigma}$ orbital.

For the cell far away from the ZR center, the local ground state $|\psi_{1h}\rangle$  with one hole filling is the slight mixing between $|d_{\uparrow}\rangle$ and  $|b_{\uparrow}\rangle$, where $|d_{\uparrow}\rangle=d^\dagger_{\uparrow} \ket{\mathrm{vac}}$ and similarly for other basis states in hole picture.  Here, we choose one spin-up polarized state without loss of generality. The lowest energy two-hole state $|\psi_{\mathrm{2h}}\rangle$ is the mixing between ZR singlet $(|d_{\uparrow}b_{\downarrow}\rangle-|d_{\downarrow}b_{\uparrow}\rangle)/\sqrt{2}$, $|b_{\uparrow}b_{\downarrow}\rangle$ and $|d_{\uparrow}d_{\downarrow}\rangle$ as an extension for ZR singlet, which can be obtained in diagonalization the block Hamiltonian
\begin{equation} \label{2hole-main}
    H^{\mathrm{2h}}=\begin{bmatrix}
        \epsilon_b+E_{\mathrm{loc}} & -2\sqrt{2}t_{pd} & -2\sqrt{2}t_{pd}  \\
        -2\sqrt{2}t_{pd} & 2\epsilon_b+2E_{\mathrm{loc}} & 0  \\
        -2\sqrt{2}t_{pd} & 0 & U \\
    \end{bmatrix}
\end{equation}

With these states, we can directly calculate the spectrum as plotted in Fig.\ref{fig2}(a). The gap $\Delta_{\mathrm{gap}}$ between $N \rightarrow N-1$ ($\omega>0$) and $N \rightarrow N+1$ ($\omega<0$) is around $2.8 ~\mathrm{eV}$ owing to the charge transfer process. 

Next, we move to the localized ZR center. The tunneling spectrum in Fig.\ref{fig2}(b) shows the gap $\Delta_{\mathrm{gap}}^{\mathrm{ZR}}$ at the ZR center has reduced to $1.2 ~\mathrm{eV}$. There is indeed an in-gap bound state inside the charge transfer $\Delta_{\mathrm{gap}}$. The essential reason for this bound state is owing to the fact that the local ground state in the doped cell has changed to $|\psi_{\mathrm{2h}}\rangle$ created by $\psi_{0A}^\dagger$ in Eq.\eqref{eq:wf}, which is extension of the ZR singlet mixed with other singlet states $|d_{\uparrow}d_{\downarrow}\rangle$ and $|b_{\uparrow}b_{\downarrow}\rangle$. The removing hole process ($N \rightarrow N-1$) is from ZR $|\psi_{\mathrm{2h}}\rangle$ to $|\psi_{1h}\rangle$ which breaks a ZR singlet, while adding hole process ($N \rightarrow N+1$) is from ZR $|\psi_{\mathrm{2h}}\rangle$  to a three-hole excitation state $|\psi_{3h}\rangle$, which are manifested by the peaks on the right and left sides of $\Delta_{\mathrm{gap}}^{\mathrm{ZR}}$, respectively.
There are two degenerate lowest energy three-hole excitation states $|\psi_{3h}\rangle=e_{1\sigma}^\dagger|\psi_{\mathrm{2h}}\rangle$ and $|\psi_{3h}\rangle=e_{2\sigma}^\dagger|\psi_{\mathrm{2h}}\rangle$.
Besides these two states, $|\psi_{3h}\rangle$ can also be other combinations leading to other peaks at lower energy. 
 It should be noted that Fig.\ref{fig2} shows the local DOS for a single CuO cell only within a low-energy range to capture the in-gap feature. A Cu/O-resolved local DOS in a broader energy range is presented in the supplementary material (SM), whose spectral features are qualitatively consistent with the cluster-DMFT results reported in Refs.\cite{kowalski2021oxygen,kumar2025cluster}. Other relevant states and tunneling processes can also be found in SM.


Clearly, the above bound state result depends on the parameters we choose. Here, we adopt the realistic parameter based on a combined local density functional theory (DFT-LDA) and quantum Monte Carlo work for doped cuprates and CCOC \cite{kent_PhysRevB.78.035132}. We also test the other parameters used in Ref. \cite{Feiner_PhysRevB.45.7959} and \cite{Chernyshev_PhysRevB.50.13768}. Similar bound states show up, although the actual gap sizes vary in different cases. 
Additionally, we also need to consider the motion effect of a single hole or a single electron in the background of quantum antiferromagnetic. 
Historically, this hole motion has been investigated by a self-consistent perturbation theory and other methods \cite{varma_PhysRevLett.60.2793,kane_PhysRevB.39.6880,Subir_PhysRevB.39.12232,Su_PhysRevLett.63.1318,Chernyshev_PhysRevB.50.13768,Yushankhai_PhysRevB.55.15562}. The resulting bandwidth $W_h$ is about $2 \sim 3$ $J$. The bandwidth $W_e$ is also around the same order \cite{Chernyshev_PhysRevB.50.13768}. Hence, the energy level we discussed above should be further broadened to $W_h$ or $W_e$, as illustrated in Fig.\ref{fig2}(c) and Fig.\ref{fig2}(d). 

\begin{figure}[t]
	\begin{center}
		\fig{3.4in}{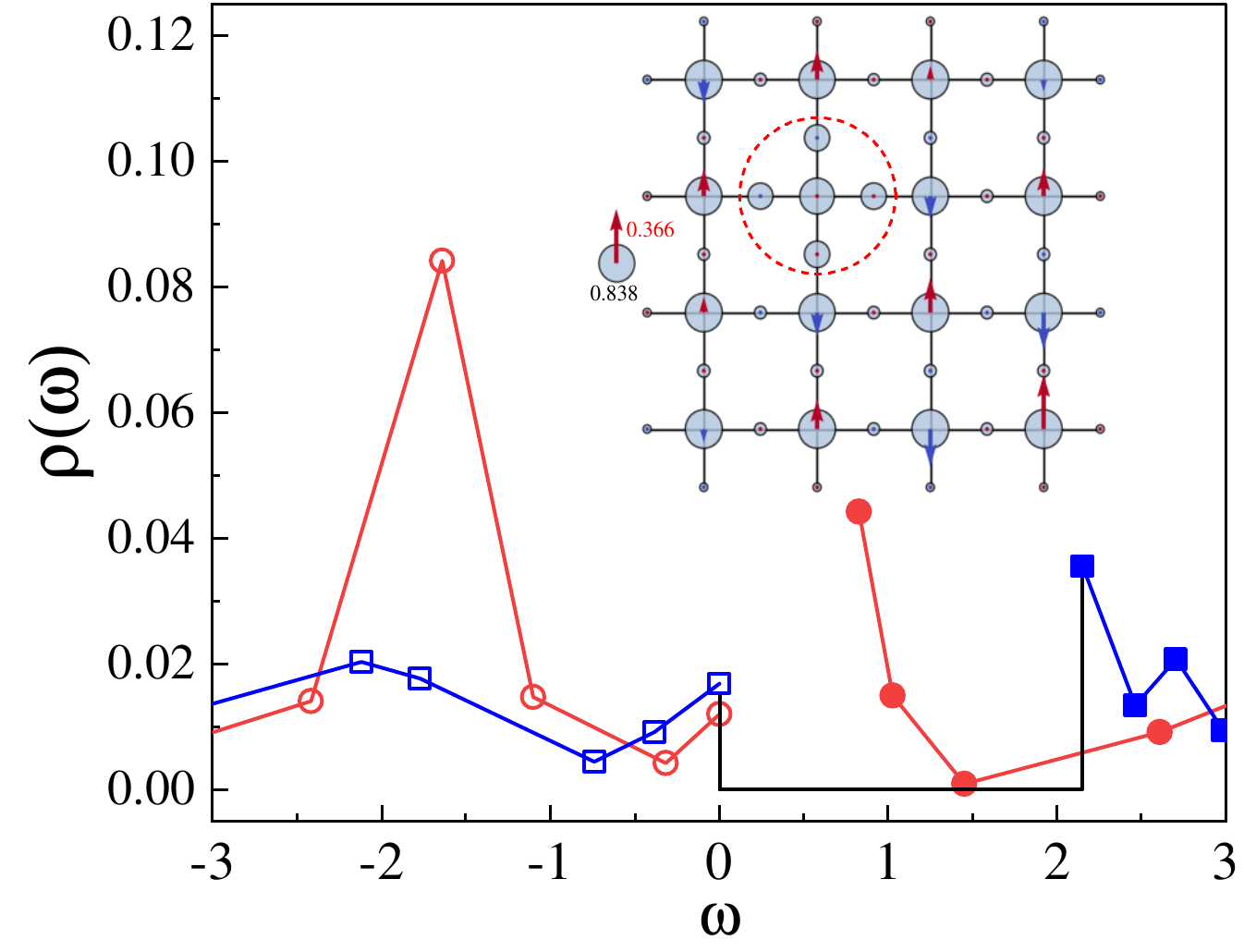}
		\caption{Local DOS $\rho(\omega)$ for a system consisted of $4 \times 4$ CuO$_2$ unit cells from MPS-based Lanczos method. Only lower energy results are plotted here and the broader energy range results can be found in SM. The blue and red lines represent results for half-filled and one extra hole-doped systems respectively. The filled (open) markers represent result for processes adding one electron (hole). The $\Delta_{\mathrm{gap}}$ for CT is highlighted by the black line.
  In the inset, we plot the spin and hole distribution for the ground state with one extra hole doping, where we can see the doped holes mainly reside on the four oxygen sites (in the dashed circle) with local energy decreased. A magnetic pinning field is also applied in the Cu site in the downright corner to show the spin pattern in the system.  The radius (length) of circles (arrows) is proportional to the hole (spin) density and the direction of the arrow means the sign of the spin density. The legend in the inset shows a site with hole density $0.838$ and spin density $0.366$.    
			\label{fig3}}
	\end{center}
\end{figure}

\subsection{MPS-based Lanczos results}

After gaining a qualitative understanding, we want to provide more evidence from many-body computation. We employ the MPS-based Lanczos method for the calculation of the spectrum function \cite{PhysRevB.85.205119}. The strategy is to represent the Lanczos basis in the matrix-product state (MPS). A special reorthogonalization process is performed after the standard Lanczos method to resolve the loss of orthogonality resulting from the compression of MPS \cite{PhysRevB.85.205119}. The spectrum functions are then obtained from the eigenstates obtained by Lehmann representation \cite{PhysRevB.85.205119}. 

The system we studied consists of $4 \times 4$ CuO$_2$ unit cells (shown in the inset of Fig.~\ref{fig3}). 
Since the state of interest in our setup is a spatially localized dopant-bound ZR-related state, with the doped hole density concentrated mainly around the impurity-centered CuO$_4$ plaquette as discussed above, the finite-size effects should be much less important here than for an extended state, and the $4 \times 4$ system is sufficient to capture the main feature.
The bond dimension for MPS in the calculation is $2000$ and the results remain almost unchanged with larger bond dimensions.  
We add the localization potential to the $4$ $p$ orbitals surrounding a Cu $d$ orbital (the dashed circle in the inset of Fig.~\ref{fig3}).
A magnetic pinning field with strength $h_m = 0.1t_{pd}$ is also applied to the Cu site in the downright corner to show the spin pattern in the system. The distribution of holes and local spin for the one hole-doped system can be found in the inset in Fig.~\ref{fig3}, where we find the doped hole is indeed pinned to the desired oxygen sites forming a localized ZR with the hole in the center Cu site.
 

The local DOS at the same Cu site for the half-filling and one-hole doped systems are plotted in Fig.~\ref{fig3}.
We distinguish the local DOS of adding electron and hole processes with filled and open markers.
For the blue square lines, the tunneling process is for the half-filling CT insulator. We obtain a charge transfer gap of about $2.15~$ eV (highlighted by the black line), which is improved closer to the experimental value than $2.8$ eV in Fig.\ref{fig2}.
More importantly, from the result of the one-hole doped case as shown in the red circle line, we can find a finite density of states inside the charge transfer gap.
This newly emerged bound state matches well with our qualitative discussion, proving the existence of localized ZR. 
The new gap $\Delta_{\mathrm{gap}}^{\mathrm{ZR}}$ is about $0.8~$ eV, which is consistent with the experimental observations \cite{ye2023visualizing}.

\subsection{Tunneling matrix effect}

For the understanding of STM measurement, we also need to take the tunneling matrix effect into account \cite{rice_tunnel,YanChen_PhysRevLett.97.237004}. STM tip coupling with the CCOC mainly through the top Cl $p_z$ orbitals as shown in Fig.\ref{fig4}(a). Hence, the operator $\phi_\sigma^\dagger(\mathbf{r})$ in Eq.\eqref{dIdV} can be modified to
\begin{eqnarray}
    \phi_\sigma^\dagger(\mathbf{r}) &=&  \sum_{\tau,i} \langle\mathbf{r}|i,\mathrm{Cl}\rangle \langle i, \mathrm{Cl}|i+\tau, O \rangle  p^\dagger_{i,\tau,\sigma} + \nonumber\\ &&\qquad \sum_{i}   \langle\mathbf{r}|i,\mathrm{Cl}\rangle \langle i, \mathrm{Cl}|i,\mathrm{ Cu} \rangle  d^\dagger_{i,\sigma} 
    \label{STM}
\end{eqnarray} 
where $\langle\mathbf{r}|i,\mathrm{Cl}\rangle$ is the overlap between STM tip at poistion $r$ and Cl $p_z$ state at site i.
$\langle i,\mathrm{Cl}|\alpha \rangle$ is the overlap between $p_z$ and other states in the CuO$_2$ plane, which inherits their wannier function information. $\tau$ labels the four O atoms nearest to the Cu atom. Notice that the overlap between $p_z$ and $d_{x^2-y^2}$ directly beneath it is strictly zero by symmetry. The first term though O $p$ becomes the dominating contribution to tunneling current. 
Since $\langle\mathbf{r}|i,\mathrm{Cl}\rangle$ strongly depends on the distance between STM tip and CCOC, we use a Gaussian function 
$|\braket{\mathbf{r}}{i,Cl}|^2\propto e^{-\frac{(\mathbf{r}-\mathbf{R}_i)^2}{L^2}}$
to evaluate the overlap, where $L$ is the characteristic length of the overlap. Fig.\ref{fig4}(b) shows the DOS map at the bound state energy. This four-lobe clover pattern matches well with the experiment observed ZR-like contour \cite{ye2023visualizing}.

\section{Discussion}

\begin{figure}[t]
	\begin{center}
		\fig{3.4in}{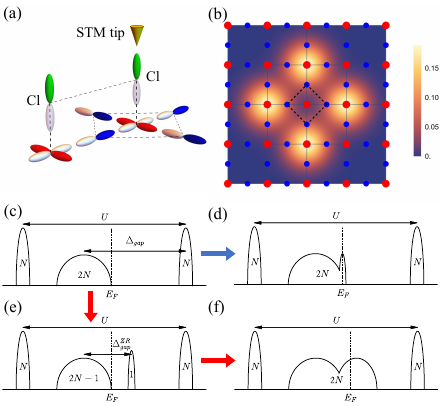}
		\caption{(a) Schematic diagram of the STM tip coupling to CCOC through $p_z$ orbitals of Cl directly above Cu $d_{x^2-y^2}$. (b) 
       Spatial STM tunneling contour of the electronic state at the energy of in-gap state on the CuO$_2$ lattice, calculated using the one-cell eigenstates together with the tunneling matrix formalism described in Eq.~\eqref{STM}. The dashed box indicates the ZR state. Here we choose L = 0.5a in the Gaussian function without loss of generality.
        (c-f) Schematic diagram of the two different scenarios of insulator-metal transition process c $\rightarrow$ d and c $\rightarrow$ e $\rightarrow$ f. (c) For the undoped CT insulator, the system is in an antiferromagnetic insulating state with a charge transfer gap. (d) In the first scenario, the doped holes emerge from the O 2p band and no in-gap states occur. (e) In the second scenario, when the system is lightly hole doped, the doped holes form localized ZR bound states emerging in the gap, and the system remains in an insulating state. (f) As the doping concentration increases, the in-gap ZR bound states begin to overlap, forming a ZR band that touches the valence band, leading to a transition to a metallic state.
			\label{fig4}}
	\end{center}
\end{figure}

A natural question arising from the in-gap bound states is how these bound states evolve into metallic states.
A straightforward calculation is challenging owing to the complicated spin pattern and random doping. 
There is one way to determine the critical hole doping density $x_c$ by comparing $E_b$ and the coherent hole motion bandwidth. We use the renormalized mean field of $t$-$J$ model based on Gutzwiller approximation to estimate bandwidth $W_{\mathrm{GW}}$ \cite{Zhang_1988,Anderson_2004}. The hopping renormalization factor is $g_t=\frac{2x}{1+x}$ while the spin exchange renormalization factor is $g_s=\frac{4}{(1+x)^2}$. $x$ is the average hole doping density $1-\langle\hat{n}\rangle$. By decoupling the $J\mathbf{S}_i\mathbf{S}_j$ into the nearest neighbor hopping bond, we find the $W_{\mathrm{GW}}$ of renormalized $t-J$ model is about 0.65 eV at $x=0.046$. Hence, the $x_c$ is estimated around $5\%$.
This estimation is consistent with experimental findings in Na-doped CCOC \cite{ccoc_0,ccoc_1,ccoc_2}, where superconductivity only emerges after $x>5\%$.
Hence, if the charge-transfer insulator is undoped, a charge-transfer gap $\Delta_{\mathrm{gap}}$ dominates the charge dynamics of the antiferromagnetic insulating ground state, as illustrated in Fig.\ref{fig4}(c).
Then, keeping light-doping the system, a bound state with the tunneling gap $\Delta_{\mathrm{gap}}^{\mathrm{ZR}}$ emerges from the charge-transfer gap as illustrated in Fig.\ref{fig4}(e), where ZR singlets are randomly localized. 
If the kinetic energy overcomes the binding energy, an itinerant ZR band overlaps with the valence band showing a metallic behavior as illustrated in Fig.\ref{fig4}(f). 

In summary, we successfully identify the origin of the in-gap bound states emerging from the lightly doped charge transfer insulator
using an approximate ground-state wavefunction based on one localized Zhang-Rice singlet and the Neel state.
By calculating the tunneling spectrum, the tunneling gap $\Delta_{\mathrm{gap}}^{\mathrm{ZR}}$ at the ZR center is around half of the charge-transfer gap $\Delta_{\mathrm{gap}}$. Moreover,
we carry out MPS-based Lanczos calculations on system consisted of $4\times4$ CuO$_2$ unit cells with local potential to pin the Zhang-Rice singlet. The results show a bound state with $0.8$ eV emerges within the charge transfer gap. Taking the STM tunneling matrix effect into account, the tunneling contour at the bounding energy around the ZR center is weak while the four neighbors are strong.
How these bound states evolve into metallic states is further discussed.
All these findings are consistent with the recent experiment observations in lightly hole-doped CCOC \cite{ye2023visualizing}.

On the other hand, we want to emphasize that the emergence of localized Zhang–Rice singlets is a generic property of doped charge-transfer insulators. The framework we introduce provides a direct and transparent means of analyzing this problem. Corresponding bound states are expected to arise in other lightly doped cuprates, such as La$_2$CuO$_4$, as well as in transition-metal compounds with a similar charge-transfer insulating character. Moreover, in addition to scanning tunneling microscopy, complementary spectroscopic probes, including optical conductivity, are capable of detecting analogous in-gap excitations. We anticipate that the present results and methodology will contribute to a broader understanding of the physics of hole-doped charge-transfer insulators.

\section{Method}
The calculations are based on the three-band Hubbard model for the CuO$_2$ plane, as defined in Eq.~\eqref{eq:three_band} \cite{Emery_PhysRevLett.58.2794,Zhang_Rice}. The active orbitals are Cu $3d_{x^2-y^2}$ and O $2p_x$, $2p_y$ orbitals, with Cu-O hopping $t_{pd}$, O-O hopping $t_{pp}$, charge-transfer energy $\tilde{\epsilon}_p-\tilde{\epsilon}_d$, and onsite interaction $U$ on Cu sites. The model parameters are taken from the ab initio estimate for CCOC used in Fig.~\ref{fig2} \cite{kent_PhysRevB.78.035132}. To model the dopant-induced localization of the Zhang-Rice singlet in the approximate ground state of Eq.~\eqref{eq:wf}, a local potential $E_{\mathrm{loc}}$ is added to the four O orbitals surrounding the doped CuO$_4$ plaquette.

For the one-cell calculation, the four O orbitals around one Cu site are decomposed into local $b_1$, $a_1$, and $e$ symmetry channels, as in Eq.~\eqref{eq:local_symmetry_main}. The resulting few-body Hamiltonians in the one-hole, two-hole, and three-hole sectors, are diagonalized exactly using the QuSpin package \cite{weinberg2017quspin,weinberg2019quspin}. The local tunneling density of states is then obtained from the matrix elements in Eq.~\eqref{dIdV} between the approximate ground state with a localized Zhang-Rice singlet and the corresponding hole-added or hole-removed excitation states. The charge gap is evaluated using Eq.~\eqref{eq:charge_gap}.

For the finite-size many-body calculation, we use an MPS-based Lanczos method to compute the local spectral function of a $4\times4$ CuO$_2$ cluster \cite{PhysRevB.85.205119}. The Lanczos basis vectors are represented as matrix-product states, and a reorthogonalization procedure is used to reduce the loss of orthogonality caused by MPS compression. The bond dimension is set to 2000, and the calculated low-energy spectral features remain stable for larger bond dimensions. A localization potential is applied to the
four O orbitals around the impurity-centered plaquette, and a weak magnetic pinning field is applied at a corner Cu site to display the antiferromagnetic spin pattern by explictly breaking the SU(2) symmetry.

\vspace{0.6cm}

\textbf{Data  availability}
The data generated in this paper are provided in the Supplementary Information/Source Data file.

\textbf{Code availability}
The code used for numerical calculation is available upon  reasonable request from the corresponding authors.

\bibliographystyle{naturemag}
\bibliography{reference}

\vspace{2cm}


\textbf{Funding:}
P.F.L., K.J. and J.P.H. acknowledge the support by the National Natural Science Foundation of China (Grant No. NSFC-12494594), the New Cornerstone Investigator Program, and the Chinese Academy of Sciences Project for Young Scientists in Basic Research (2022YSBR-048). T.X. acknowledges the support by the National Natural Science Foundation of China (Grant No. 12488201), the National Key Research and Development Program of China (2021ZD0301800). Y.S. and M.P.Q. acknowledge the support by the Ministry of Science and Technology (Grant No. 2022YFA1405400), the National Natural Science Foundation of China (NSFC-12274290 and NSFC-12522406) and the Innovation Program for Quantum Science and Technology (2021ZD0301902).

\textbf{Author contributions}: P.F.L., K.J., Y.S. and M.P.Q. performed the theoretical and numerical calculations. J.P.H. and T.X. provided the theoretical understanding. K.J. conceived the work. All authors analyzed the results, wrote and reviewed the manuscript.

\textbf{Competing interests}: The authors declare no competing interests.

\clearpage
\onecolumngrid
\begin{center}
\textbf{\large Supplementary Information: Bound states in doped charge transfer insulators}
\end{center}

\setcounter{equation}{0}
\setcounter{figure}{0}
\setcounter{table}{0}
\setcounter{page}{1}
\makeatletter
\renewcommand{\theequation}{S\arabic{equation}}
\renewcommand{\thefigure}{S\arabic{figure}}
\renewcommand{\thetable}{S\arabic{table}}
\newcommand{\dg}{{\dagger}}
\newcommand{\upa}{{\uparrow}}
\newcommand{\dna}{{\downarrow}}
\newcommand{\br}{{\mathbf{r}}}

\onecolumngrid

\subsection{LDOS for broader energy range for the $4 \times 4$ system}
In Supplementary Fig.~\ref{LDOS-full}, we show the LDOS result for the same system as in Fig.~3 in the main text (consisted of $4 \times 4$ CuO$_2$ unit cells) from MPS-based Lanczos method. In Fig.~3 in the main text, results between $-3$ eV and $3$ eV are shown, while in Supplementary Fig.~\ref{LDOS-full}, the LDOS for broader energy range is plotted, where we can identify the lower (around $-9$ eV) and upper (around $4$ eV) Hubbard bands.  
\begin{figure}[!htbp]
	\begin{center}
		\fig{3.4in}{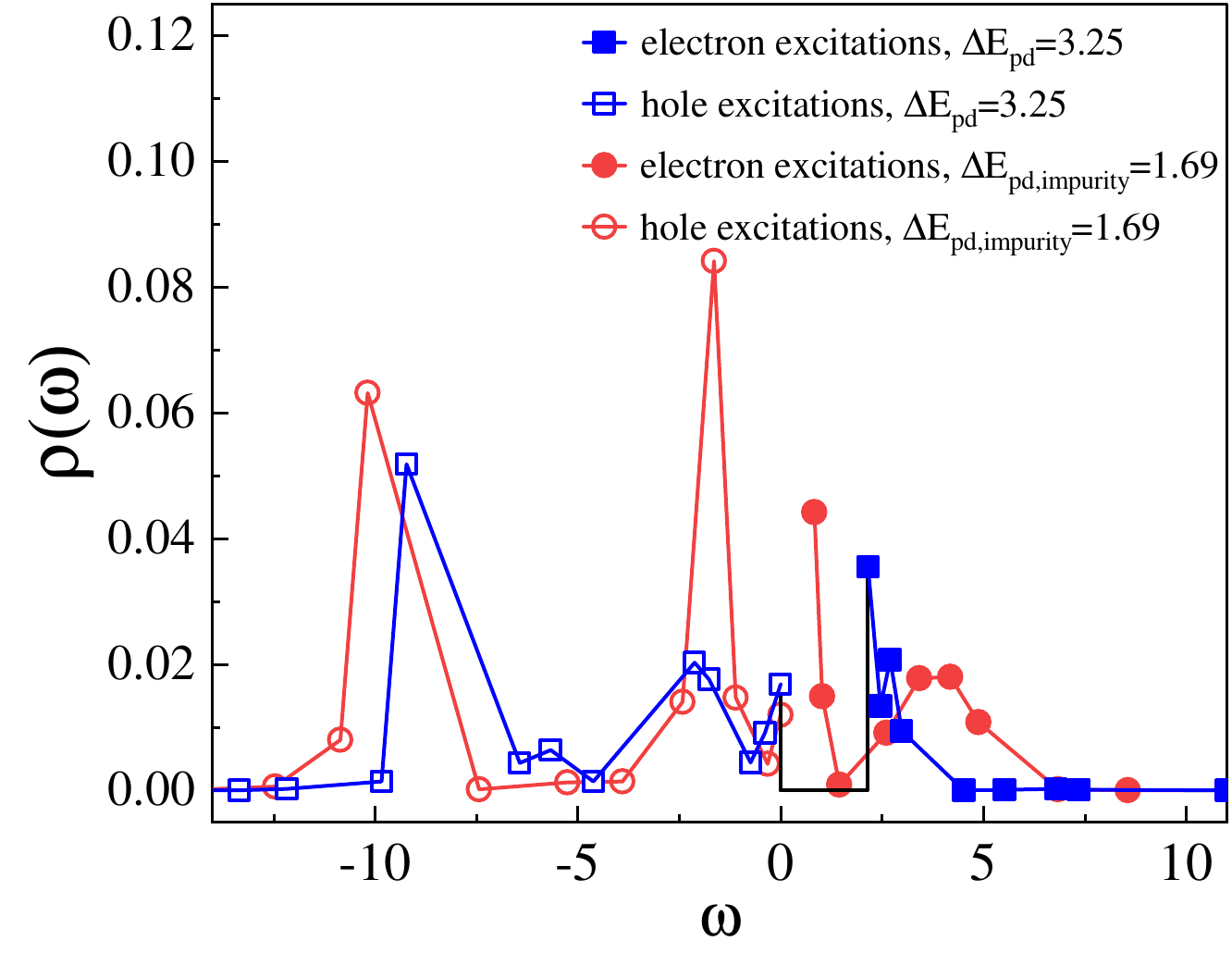}
		\caption{LDOS for the same system in Fig.~3 in the main text (consisted of $4 \times 4$ CuO$_2$ unit cells) from MPS-based Lanczos method. Here the energy range is broader than that in Fig.~3.      
			\label{LDOS-full}}
	\end{center}
\end{figure}

\vskip -0.5cm
\onecolumngrid
\subsection{Relevant states in one-cell calculation}
In this section, we list all the eigenstates relevant to our discussion and the STM spectrum. 
The general strategy for our approach is that we can use one-cell eigenstates to approximate the eigenstates of localized ZR singlets. Following the three-band Hamiltonian, the local operators are Cu 3$d$ and O 2$p$ orbitals.
Using the local $D_4$ symmetry, we can group the four $p$ orbital into one $b_1$, one $a_1$ and two degenerate $e$ symmetric basis:
\begin{eqnarray}
    b_{\sigma}&=&(p_{x\sigma}-p_{y\sigma}-p_{-x\sigma}+p_{-y\sigma})/2 \nonumber \\
    a_{\sigma}&=&(p_{x\sigma}+p_{y\sigma}-p_{-x\sigma}-p_{-y\sigma})/2  \label{eqp} \\
    e_{1\sigma}&=&(p_{x\sigma}+p_{-x\sigma})/{\sqrt{2}}, e_{2\sigma}=(p_{y\sigma}+p_{-y\sigma})/{\sqrt{2}}  \nonumber 
\end{eqnarray}
Their corresponding energies become $\epsilon_b=\tilde{\epsilon}_p-2t_{pp}$,$\epsilon_a=\tilde{\epsilon}_p+2t_{pp}$ and $\epsilon_e=\tilde{\epsilon}_p$. Due to $d_\sigma$ only has $B_1$ symmetry, it only has overlap with $b_{\sigma}$ orbital.
The one-cell Hamiltonian can be written as
\begin{eqnarray}
    H_0&=& \sum_{\sigma} \epsilon_b b_{\sigma}^\dagger b_{\sigma}+\epsilon_a a_{\sigma}^\dagger a_{\sigma}+\epsilon_e (e_{1\sigma}^\dagger e_{1\sigma}+e_{2\sigma}^\dagger e_{2\sigma}) \nonumber \\
    &-&2t_{pd} (b_{\sigma}^\dagger d_{\sigma}+d_{\sigma}^\dagger b_{\sigma} )+\epsilon_d d_{\sigma}^\dagger d_{\sigma}+U n_{\uparrow}^{d}  n_{\downarrow}^{d} \nonumber \\
    &+&\sum_{\sigma} E_{\mathrm{loc}} (b_{\sigma}^\dagger b_{\sigma}+a_{\sigma}^\dagger a_{\sigma}+e_{1\sigma}^\dagger e_{1\sigma}+e_{2\sigma}^\dagger e_{2\sigma})
\end{eqnarray}
where the $n_{\sigma}^{d}$ is the density operator for $d$ electrons. The vacuum $|0\rangle$ is always defined for fully occupied Cu 3$d^{10}$ and O 2$p^{6}$. And we adopt the notation in Ref. \cite{Feiner_PhysRevB.45.7959}. We briefly use $|\alpha \rangle$ for $\alpha^\dagger|0\rangle$ in the main text and the following discussion.
Then, we can represent the many-body Fock states and map the $H_0$ into each number hole space.

(1) \textit{1-hole}
\begin{table}[]
    \centering
    \begin{tabular}{c|c|c}
    \hline
       Symmetry  & $|n\rangle$ & E$_n$  \\
    \hline
        $B_1$     & $\ket{S_0^{1|B_1}}$  & -1.013 eV  \\
        $B_1$     & $\ket{S_1^{1|B_1}}$  & 2.403 eV  \\
        $A_1$     & $|S^{1|A_1}\rangle$  & 1.99 eV \\
        $E$       & $|S^{1|E}\rangle$  & 1.69 eV \\
    \hline
    \end{tabular}
    \caption{One hole states and their eigenvalues. Superscripts of $\ket{S}$ denote the symmetry representation to which the state belongs, subscripts indicate the ordering of eigenvalues of the block Hamiltonian, and a subscript of 0 represents the lowest energy state.}
        \label{tab:1-hole}
\end{table}

For the undoped case, there is one hole in the cell ($3d^{9}2p^{6}$, $3d^{10}2p^{5}$ configurations). 
The many-body Fock states can be shown as $\ket{d_\sigma}$, $\ket{b_\sigma}$, $\ket{a_\sigma}$ and $\ket{e_{1\sigma}}$, $\ket{e_{2\sigma}}$. 
It is useful to split Hamiltonian into different symmetry representations under $C_{4v}$ and spin SU(2). The results are listed in Supplementary Table~\ref{tab:1-hole}. The lowest energy belongs to the B$_1$ sector. 
\begin{itemize}
\item B$_1$: Since the hopping term $t_{pd}$ in the Hamiltonian only mixes two states $\ket{d_\sigma}$ and $\ket{b_\sigma}$, they gain the energy of hopping and the block Hamiltonian in $\ket{d_\sigma}$, $\ket{b_\sigma}$ basis is 
\begin{equation}
    h^{1|B_1}=\begin{bmatrix}
        0 & -2t_{pd} \\
        -2t_{pd} & \epsilon_b+E_{\mathrm{loc}}
    \end{bmatrix}
\end{equation}
where $E_{\mathrm{loc}}$ is the local potential energy due to the attraction to the doped hole from the dopant. The ground state of the Hamiltonian 
\begin{equation}
    \ket{S_0^{1|B_1}}=0.839\ket{d_\sigma}+0.545\ket{b_\sigma}
\end{equation}
with energy $ E_0^{1|B_1}=-1.013~\rm{eV}$ is called the bonding state and the other eigenstate is the anti-bonding state $\ket{S_1^{1|B_1}}$ with very high energy $ E_1^{1|B_1}=2.403~\rm{eV}$. 
\item A$_1$: $\ket{S^{1|A_1}}=\ket{a_\sigma}$ with energy $\epsilon_a+E_{\mathrm{loc}}=1.99$~eV.
\item E: $\ket{S^{1|E}}=\ket{e_{1/2,\sigma}}$ with energy $\epsilon_e+E_{\mathrm{loc}}=1.69$~eV.
\end{itemize}

\begin{figure}[t] 
	\begin{center}
		\fig{5.4in}{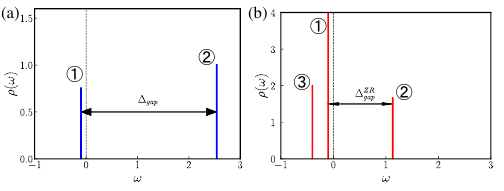}
		\caption{Tunneling DOS spectrum calculated away from the ZR center and at the ZR center of ground state $\ket{\mathrm{GS}}$ respectively.}
            \label{peak}
	\end{center}
\end{figure}

(2) \textit{2-hole}
\begin{table}[]
    \centering
    \begin{tabular}{c|c|c}
    \hline
       Symmetry  & $|n\rangle$ & E$_n$  \\
    \hline
        singlet $A_1$     & $\ket{S_0^{2|A_1}}$  & -0.554 eV  \\
        singlet $A_1$     & $\ket{S_1^{2|A_1}}$  & 4.119 eV  \\
        singlet $A_1$     & $\ket{S_2^{2|A_1}}$  & 10.105 eV \\
        $B_1$     & $\ket{S_0^{2|B_1}}$  &  1.318 eV \\
        $B_1$     & $\ket{S_1^{2|B_1}}$  & 5.612 eV \\
        $E$     & $\ket{S_0^{2|E}}$  &  0.677 eV \\
        $E$     & $\ket{S_1^{2|E}}$  & 4.093 eV \\
    \hline
    \end{tabular}
    \caption{Two hole states and their eigenvalues}
    \label{tab:2-hole}
\end{table}

If a hole is doped into the cell, we need to analyze a two-hole system ($3d^{8}2p^{6}$, $3d^{9}2p^{5}$, $3d^{10}2p^{4}$ configurations). The results are listed in Supplementary Table~\ref{tab:2-hole}. The lowest energy belongs to the spin-singlet $A_1$ sector.
\begin{itemize}
\item 
Spin-singlet $A_1$: We can write the block Hamiltonian under the basis states $(\ket{d_\uparrow b_\downarrow}-\ket{d_\downarrow b_\uparrow})/\sqrt{2}$, $\ket{b_\upa b_\dna}$ and $\ket{d_\upa d_\dna}$ as
\begin{equation} \label{2hole}
    h^{2|A_1}=\begin{bmatrix}
        \epsilon_b+E_{\mathrm{loc}} & -2\sqrt{2}t_{pd} & -2\sqrt{2}t_{pd}  \\
        -2\sqrt{2}t_{pd} & 2\epsilon_b+2E_{\mathrm{loc}} & 0  \\
        -2\sqrt{2}t_{pd} & 0 & U 
    \end{bmatrix}
\end{equation}
 Note that we only consider the basis states which are coupled by $t_{pd}$, so we omit the $\ket{a_\upa a_\dna}$ here. The basis function $(\ket{d_\uparrow b_\downarrow}-\ket{d_\downarrow b_\uparrow})/\sqrt{2}$ is the just ZR singlet in their original paper \cite{Zhang_Rice}. If we use the parameters as in the main text, the ground state of the block Hamiltonian is 
\begin{equation}
    \ket{S_0^{2|A_1}}=\frac{0.820}{\sqrt{2}}(\ket{d_\uparrow b_\downarrow}-\ket{d_\downarrow b_\uparrow})+0.543\ket{b_\upa b_\dna}+0.180\ket{d_\upa d_\dna}
\end{equation}
with energy $E_0^{2|A_1}=-0.554~\rm{eV}$. Thus, the main part of the ground state is still ZR singlet. Other eigenvalues of $h^{2|A_1}$ are listed in Supplementary Table.\ref{tab:2-hole}.

\item Spin-triplet $A_1$:
The basis states $(\ket{d_\uparrow b_\downarrow}+\ket{d_\downarrow b_\uparrow})/\sqrt{2}$, $\ket{d_\upa b_\upa}$ and $\ket{d_\dna b_\dna}$ are decoupled. The energy of them are all $\epsilon_b+E_{\mathrm{loc}}=1.39 $~eV.

\item
$B_1$: The block Hamiltonian under the basis states $\ket{d_\sigma a_\sigma}$ and $\ket{b_\sigma a_\sigma}$ is
\begin{equation} \label{2hole}
    h^{2|B_1}=\begin{bmatrix}
        \epsilon_a+E_{\mathrm{loc}} & -2t_{pd}  \\
        -2t_{pd} & \epsilon_b+\epsilon_a+2E_{\mathrm{loc}} 
    \end{bmatrix}
\end{equation}
We can find that $h^{2|B_1}=h^{1|B_1}+\epsilon_a+E_{\mathrm{loc}}$. 
\item
$E$:  The block Hamiltonian under basis $\ket{d_\sigma e_\sigma}$ and $\ket{b_\sigma e_\sigma}$ is $h^{2|E}=h^{1|B_1}+\epsilon_e+E_{\mathrm{loc}}$ with the eigenvalues $E^{2|E}=E^{1|B_1}+\epsilon_e+E_{\mathrm{loc}}$.
\item 
Other sectors are not coupled by $t_{pd}$ and the energy of them are all higher than $E_0^{2|A_1}$.

\end{itemize}

(3) \textit{3-hole}

\begin{table}[]
    \centering
    \begin{tabular}{c|c|c}
    \hline
       Symmetry  & $|n\rangle$ & $E_n$  \\
    \hline
        $E$     & $\ket{S_0^{3|E}}$  & 1.136 eV  \\
        $E$     & $\ket{S_1^{3|E}}$  & 5.809 eV  \\
        $E$     & $\ket{S_2^{3|E}}$  & 11.795 eV  \\
        $B_1$     & $\ket{S_0^{3|B_1}}$  &  2.49 eV \\
        $B_1$     & $\ket{S_1^{3|B_1}}$  & 11.18 eV  \\
        $A_1$     & $\ket{S_0^{3|A_1}}$  & 1.436 eV  \\
        $A_1$     & $\ket{S_1^{3|A_1}}$  & 6.109 eV  \\
        $A_1$     & $\ket{S_2^{3|A_1}}$  & 12.095 eV  \\
    \hline
    \end{tabular}
    \caption{Three hole states and their eigenvalues}
    \label{tab:3-hole}
\end{table}

For 3-hole state ($3d^{7}2p^{6}$, $3d^{8}2p^{5}$, $3d^{9}2p^{4}$, $3d^{10}2p^{3}$ configurations), the lowest-energy state is located in the $E$ representation $S_z=\pm1/2$ subspace. The results are listed in  Supplementary Table.\ref{tab:3-hole}.

\begin{itemize}
\item 
$S=1/2$, $S_z=1/2$, $E$: With basis $\ket{d_\upa d_\dna e_\upa}$, $\frac{1}{\sqrt{2}}\left(\ket{d_\upa b_\dna e_\upa} - \ket{d_\dna b_\upa e_\upa}\right)$ and $\ket{b_\upa b_\dna e_\upa}$, the block Hamiltonian is
\begin{equation}
    h^{3|E}=\begin{bmatrix}
        \epsilon_e+U+E_{\mathrm{loc}} & -2\sqrt{2}t_{pd} & 0 \\
        -2\sqrt{2}t_{pd} & \epsilon_b+\epsilon_e+2E_{\mathrm{loc}} & -2\sqrt{2}t_{pd}\\
        0 & -2\sqrt{2}t_{pd} & 2\epsilon_b+\epsilon_e+3E_{\mathrm{loc}}
    \end{bmatrix}
\end{equation}
The lowest energy eigenstate is 
\begin{equation}
    \ket{S_0^{3|E}}=0.180\ket{d_\upa d_\dna e_\upa} + \frac{0.820}{\sqrt{2}}\left(\ket{d_\upa b_\dna e_\upa} - \ket{d_\dna b_\upa e_\upa}\right) + 0.543\ket{b_\upa b_\dna e_\upa}
\end{equation} 
with eigenvalue $E_0^{3|E}=1.136~\rm{eV}$. We can find $\ket{S_0^{3|E}}=e_\sigma^\dg\ket{S_0^{2|A_1}}$.

\item 
$S=1/2$, $S_z=1/2$, $B_1$: The block Hamiltonian under the basis states $\ket{d_\upa b_\upa d_\dna }$ and $\ket{d_\upa b_\upa b_\dna }$ is
\begin{equation} \label{2hole}
    h^{3|B_1}=\begin{bmatrix}
        \epsilon_b+E_{\mathrm{loc}}+U & -2t_{pd}  \\
        -2t_{pd} & 2\epsilon_b+2E_{\mathrm{loc}} 
    \end{bmatrix}
\end{equation}
The lowest eigenvalue of $h^{3|B_1}$ is $E_0^{3|B_1}=2.49$~eV which is higher than $E_0^{3|E}$.

\item 
$S=1/2$, $S_z=1/2$, $A_1$: The block Hamiltonian  under the basis states$(\ket{d_\uparrow b_\downarrow a_\upa}-\ket{d_\downarrow b_\uparrow a_\upa})/\sqrt{2}$, $\ket{b_\upa b_\dna a_\upa}$ and $\ket{d_\upa d_\dna a_\upa}$ is just $h^{3|A_1}=h^{2|A_1}+\epsilon_a+E_{\mathrm{loc}}$
The lowest eigenvalue of $h^{3|A_1}$ is $E_0^{3|A_1}=1.436$~eV which is also higher than $E_0^{3|E}$.

\item 
Other sectors are irrelevant to our discussion and energies of them are all higher than $E_0^{3|E}$.

\end{itemize}

\textit{Tunneling process}
With all this information, we can identify the tunneling spectrum in Supplementary Fig.\ref{peak} 
shows the DOS spectrum away from the ZR center and at the ZR center respectively. For the cell away from the ZR center, there's no hole-doped in it, so the ground state is the 1-hole state $\ket{S_0^{1|B_1}}$ but without $E_{\mathrm{loc}}$.  The lowest-energy hole excitation is $|S_0^{1|B_1}\rangle \rightarrow \ket{S_0^{2|A_1}}$ showed by peak \ding{172} in Supplementary Fig.\ref{peak} (a). The particle excitation manifested by peak \ding{173} is the $|S_0^{1|B_1}\rangle \rightarrow |0\rangle$ state. Since there's no local potential far away from the ZR center, the energy $E_{0}^{1|B_1}=-0.672~\rm{eV}$ and $E_0^{2|A_1}=1.299~\rm{eV}$ are calculated by setting $E_{\mathrm{loc}}=0$. Thus, the energy gap observed in STM spectrum $\Delta_{\rm{gap}}=0+E_0^{2|A_1}-2E_0^{1|B_1}=2.643~\rm{eV}$ which is just the Mott gap.

For the cell at the ZR center in Supplementary Fig.\ref{peak} (b), the ground state becomes $\ket{S_0^{2|A_1}}$. The peak \ding{172} manifests the lowest hole excitation $\ket{S_0^{2|A_1}} \rightarrow \ket{S_0^{3|E}}$, while the peak \ding{173} is the lowest particle excitation $\ket{S_0^{2|A_1}} \rightarrow \ket{S_0^{1|B_1}}$ which is the state breaking a ZR singlet. Thus, the energy gap $\Delta^{\mathrm{ZR}}_{\rm{gap}}=E_0^{3|E}+E_0^{1|B_1}-2E_0^{2|A_1}=1.231~\rm{eV}$ describes a in-gap state. The peak \ding{174} is the hole excitation state  with higher energy $\ket{S_0^{2|A_1}} \rightarrow \ket{S_0^{3|A_1}}$. Actually, if we zoom in the energy window, there exist more high-energy excitations which are not relevant with the in-gap state.

\subsection{Cu/O-resolved LDOS in a broader energy range}

\begin{figure}[t]
    \centering
    \includegraphics[width=0.72\linewidth]{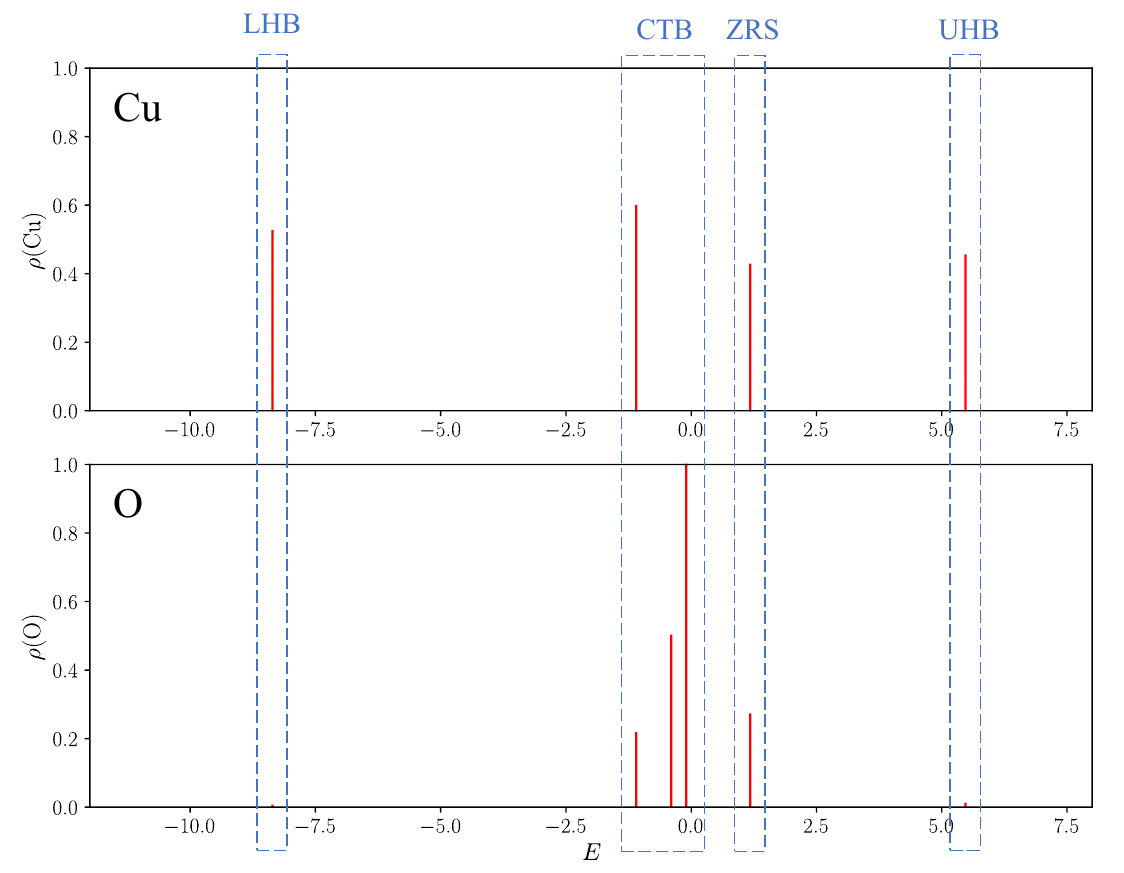}
    \caption{LDOS for a single CuO cell at ZR center over a broader energy range, resolved into Cu and O contributions in the top panel and bottom panel respectively. }
    \label{fig:CuO}
\end{figure}

To complement the low-energy LDOS shown in Fig.~2 in the main text, here we present in Supplementary Fig.~\ref{fig:CuO} the LDOS over a broader energy range, resolved into Cu and O contributions for a single CuO cell at the ZR center. The Hamiltonians in the one-hole, two-hole, and three-hole sectors, are diagonalized exactly using the QuSpin package \cite{weinberg2017quspin,weinberg2019quspin}.

As shown in Supplementary Fig.~\ref{fig:CuO}, the spectral structure contains the characteristic features commonly discussed for correlated three-band descriptions of cuprates, namely the lower-Hubbard-like feature (LHB), the charge-transfer band (CTB), the Zhang--Rice-like state (ZRS), and the upper Hubbard band (UHB). The Cu-resolved spectrum carries the dominant weight in the UHB region and also visible weight in the lower-energy Hubbard sector, whereas the O-resolved spectrum is strongest in the CTB region. In the ZRS region, both Cu and O components are present, indicating the mixed Cu--O character expected for a Zhang--Rice-like state.

Therefore, the spectral features of Supplementary Fig.~\ref{fig:CuO} are qualitatively consistent with the cluster-DMFT results reported in Refs.\cite{kowalski2021oxygen,kumar2025cluster}, where the ZRS-related low-energy state is found to have mixed Cu--O character, the CTB is predominantly oxygen-derived, and the UHB is mainly Cu-like. In our finite-cell calculation, these structures appear as groups of discrete many-body levels rather than smooth bands, so weak admixtures may be distributed among nearby levels instead of appearing as a broad continuum tail. Nevertheless, the overall orbital character and spectral ordering remain the same at the qualitative level.

\subsection{Spatial STM map of the in-gap state}
In this section, we want to provide more details on the spatial STM map in Fig.~4(b) in the main text.
Experimentally, any STM maps come from their tunneling currents. For CCOC, the top surface layer is the Cl layer\cite{YanChen_PhysRevLett.97.237004}. Therefore, most STM tunneling currents are through the Cl p$_z$ orbital. The hopping between Cl p$_z$ orbital and Cu 3d, O 2p orbitals (or their corresponding Wannier orbitals) can be easily constructed based on slater-koster formalism.
Mathematically, the spatial STM map can be evaluated by Eq.3
\begin{eqnarray}
    \frac{dI}{dV}(\mathbf{r},\omega) \propto \sum_{n} |\langle n| \phi_{\sigma}^{\dagger}(\mathbf{r}) |GS\rangle|^2 \delta(-\omega-E_n+E_{\mathrm{GS}}) 
    +\sum_{m} |\langle m| \phi_{\sigma}(\mathbf{r})|GS\rangle|^2 \delta(-\omega-E_{\mathrm{GS}}+E_{m})
\end{eqnarray} 
where $E_{\mathrm{GS}}=E_0^{2|A_1}$ is the ground state of hole-doped CuO cell.
All information is reflected in the operator $\phi_\sigma^\dg(\br)$ in Eq.3, which is the electron creation operator through Cl atom (including the nearest and second nearest neighbors) as shown by Eq.(6)
\begin{align}
    \phi_\sigma^\dagger(\mathbf{r}) &=  \sum_{\tau,i} \braket{\mathbf{r}}{i,Cl} \langle i, Cl|i+\tau, O \rangle  p^\dagger_{i,\tau,\sigma} + \sum_{i}   \braket{\mathbf{r}}{i,Cl} \langle i, Cl|i, Cu \rangle  d^\dagger_{i,\sigma} 
\end{align}
where $p^\dagger_{i,\tau,\sigma}$ is the creation operator of the local $p_x$ or $p_y$ orbital of O atom around Cu at site $i$ and $d^\dagger_{i,\sigma}$ is the creation operator of the local $d_{x^2-y^2}$ orbital of Cu atom at the site $i$. Since Cu $d_{x^2-y^2}$ and Cl $p_z$ electron overlapping is zero by symmetry, only the first terms are important to our tunneling spectrum. Then,
$\phi_\sigma^\dg(\br)$ becomes
\begin{align}
    \phi_\sigma^\dagger(\mathbf{r}) &=  \sum_{\tau,i} (-1)^{P_\tau} \braket{\mathbf{r}}{i,Cl} p^\dagger_{i,\tau,\sigma}
\end{align}
The coupling between the STM tip at position $r$ and Cl p$_z$ orbital at position $i$ becomes the most important part. Here, we use a Gaussian function to evaluate this overlap as 
\begin{equation}
    |\braket{\mathbf{r}}{i,Cl}|^2\propto e^{-\frac{(\br-\mathbf{R}_i)^2}{L^2}}
\end{equation}
The Gaussian parameter $L$ depends on the distance between STM tip and Cl plane. Here we choose $L=0.5a$ without loss of generality.

What we concern about is the spatial STM map of the in-gap state at $E_n=E_0^{1|B_1}$ and thus $\ket{n}=\ket{S_0^{1|B_1}}$. Since only the doped cell has the excited state with energy $E^{1|B_1}$, $\phi_\sigma^\dagger(\mathbf{r})$ can be expanded as
\begin{equation}
\phi_\sigma^\dagger(\mathbf{r})=\langle\br|R,Cl\rangle(p_{x\sigma}^\dagger+p_{y\sigma}^\dagger-p_{-x\sigma}^\dagger-p_{-y\sigma}^\dagger)/2+\sum_\tau(-1)^{P_\tau}\langle\br|R+\tau,Cl\rangle p_{\tau\sigma}^\dagger
\end{equation}
where $R$ denotes the ZR center. Obviously, the first term is just $a_1^\dagger$ of O atom. In our one-cell calculation, $a_1$ orbital is not a component of the in-gap state. Hence, the first term is zero. However, the matrix element in the second term is  $|\bra{S_0^{1|B_1}}p^\dg_{\tau\sigma}\ket{S_0^{2|A_1}}|^2=0.306$.

\end{document}